\documentclass[aps,reprint,superscriptaddress,shortbibliography,prl]{revtex4-1}

\usepackage{graphicx, amsmath, verbatim, dsfont, amsfonts, color}
\usepackage{braket}

\usepackage[us]{datetime}

\begin{document}

\title{Tuning insulator-semimetal transitions in 3D topological insulator thin films by inter-surface hybridization and in-plane magnetic fields}

\author{Yang~Xu}
\affiliation{Department of Physics and Astronomy, Purdue University, West Lafayette, IN 47907, USA}
\affiliation{Birck Nanotechnology Center, Purdue University, West Lafayette, IN 47907, USA}
\author{Guodong~Jiang}
\affiliation{Department of Physics and Astronomy, Purdue University, West Lafayette, IN 47907, USA}
\author{Ireneusz~Miotkowski}
\affiliation{Department of Physics and Astronomy, Purdue University, West Lafayette, IN 47907, USA}
\author{Rudro~R.~Biswas}
\affiliation{Department of Physics and Astronomy, Purdue University, West Lafayette, IN 47907, USA}
\affiliation{Purdue Quantum Center, Purdue University, West Lafayette, IN 47907, USA}
\author{Yong~P.~Chen}
\email{yongchen@purdue.edu}
\affiliation{Department of Physics and Astronomy, Purdue University, West Lafayette, IN 47907, USA}
\affiliation{Birck Nanotechnology Center, Purdue University, West Lafayette, IN 47907, USA}
\affiliation{Purdue Quantum Center, Purdue University, West Lafayette, IN 47907, USA}
\affiliation{School of Electrical and Computer Engineering, Purdue University, West Lafayette, IN 47907, USA}
\affiliation{WPI-AIMR International Research Center on Materials Sciences, Tohoku University, Sendai, 980-8577 Japan}
\date{\today}

\begin{abstract}

 A pair of Dirac points (analogous to a vortex-antivortex pair) associated with opposite topological numbers (with $\pm\pi$ Berry phases) can be merged together through parameter tuning and annihilated to gap the Dirac spectrum, offering a canonical example of a topological phase transition. Here, we report transport studies on thin films of BiSbTeSe$_2$ (BSTS), which is a 3D TI that hosts spin-helical gapless (semi-metallic) Dirac fermion surface states (SS) for sufficiently thick samples, with an observed resistivity close to $h/4e^2$ at the charge neutral point. When the sample thickness is reduced to $\sim$10 nm thick, the Dirac cones from the top and bottom surfaces can hybridize (analogous to a ``merging'' in the real space) and become gapped to give a trivial insulator. Furthermore, we observe that an in-plane magnetic field can drive the system again towards a metallic behavior, with a prominent negative magnetoresistance (MR, up to $\sim$$-$95\%) and a temperature-insensitive resistivity close to $h/2e^2$ at the charge neutral point. The observation is interpreted in terms of a predicted effect of an in-plane magnetic field to reduce the hybridization gap (which, if small enough, may be smeared by disorder and a metallic behavior). A sufficiently strong magnetic field is predicted to restore and split again the Dirac points in the momentum space, inducing a distinct 2D topological semimetal (TSM) phase with 2 single-fold Dirac cones of opposite spin-momentum windings. 
\end{abstract}

\maketitle

A wide range of quantum materials including graphene, topological insulators (TIs), Dirac/Weyl semimetals, and their artificial analogues, have been identified whose low-energy excitations behave as massless Dirac particles to host novel relativistic quantum phenomena \cite{CastroNeto2009,Hasan2010,Qi2011,Armitage2018,Lu2014,Bellec2013,Tarruell2012}. The Dirac spectra can be gapped by breaking the underlying symmetry that protects the Dirac points (DPs), or by pairwise merging and annihilation of DPs \cite{Bellec2013,Tarruell2012,Wehling2014,Hasegawa2006,Baik2015,Pereira2009,Feilhauer2015}. Previously predicted material platforms to explore this latter mechanism, such as graphene with engineered anisotropic nearest-neighbor hopping \cite{Hasegawa2006} and thin black phosphorus under a strong electric field \cite{Baik2015}, require extreme parameter tuning that is difficult to realize experimentally \cite{Pereira2009,Feilhauer2015,Kim2015}. Alternative platforms that have enabled experimental demonstration of this effect include a microwave analogue of strained graphene \cite{Bellec2013} and cold atoms in honeycomb optical lattices \cite{Tarruell2012}. On the other hand, 3D TI thin films with hybridization gapped surface states bring new opportunities to study such topological transitions in a solid-state system. In particular, merging and annihilating of top and bottom surface DPs (with opposite spin windings) can be controlled both in the real space (by sample thickness, for example demonstrated experimentally in Ref. \cite{Zhang2010} by angle-resolved photoemission spectroscopy on thin films grown by molecular beam epitaxy) and the momentum space (by an in-plane magnetic field, as theoretically proposed in Ref. \cite{Zyuzin2011a}).

In a relatively thick 3D TI film (thickness $t\gg10$ nm), the top and bottom surfaces are well separated and their corresponding topological SS Dirac cones are gapless with opposite spin helicities. When the sample is thin enough (typically $\leq\sim$10 nm) to enable hybridization between the two surfaces, a gap $\Delta_0$ is opened at the DP (even though the time-reversal symmetry is still preserved). The SS band structure acquires massive Dirac dispersion $\varepsilon=\pm\sqrt{(\hbar v_f k)^2+(\Delta_0 /2)^2}$, with $\hbar$ being the Plank constant $h$ divided by $2\pi$, $v_f$ the Fermi velocity and $k$ the (in-plane) wave vector \cite{Zhang2010}. Such a crossover of 3D TIs to the two-dimensional (2D) limit, as well as their response to magnetic fields, is little explored by electronic transport measurements in 3D TI materials with no bulk conduction and surface-dominant transport (such as BSTS). Previous in-plane magneto-transport studies in 3D TIs often suffer from their residual bulk conduction \cite{Wiedmann2016,Breunig2017} and few have been reported in the hybridization regime \cite{Lin2013,Liao2015,Taskin2017}.

\begin{figure*}\label{fig:1}
	\centering\includegraphics[width=2\columnwidth]{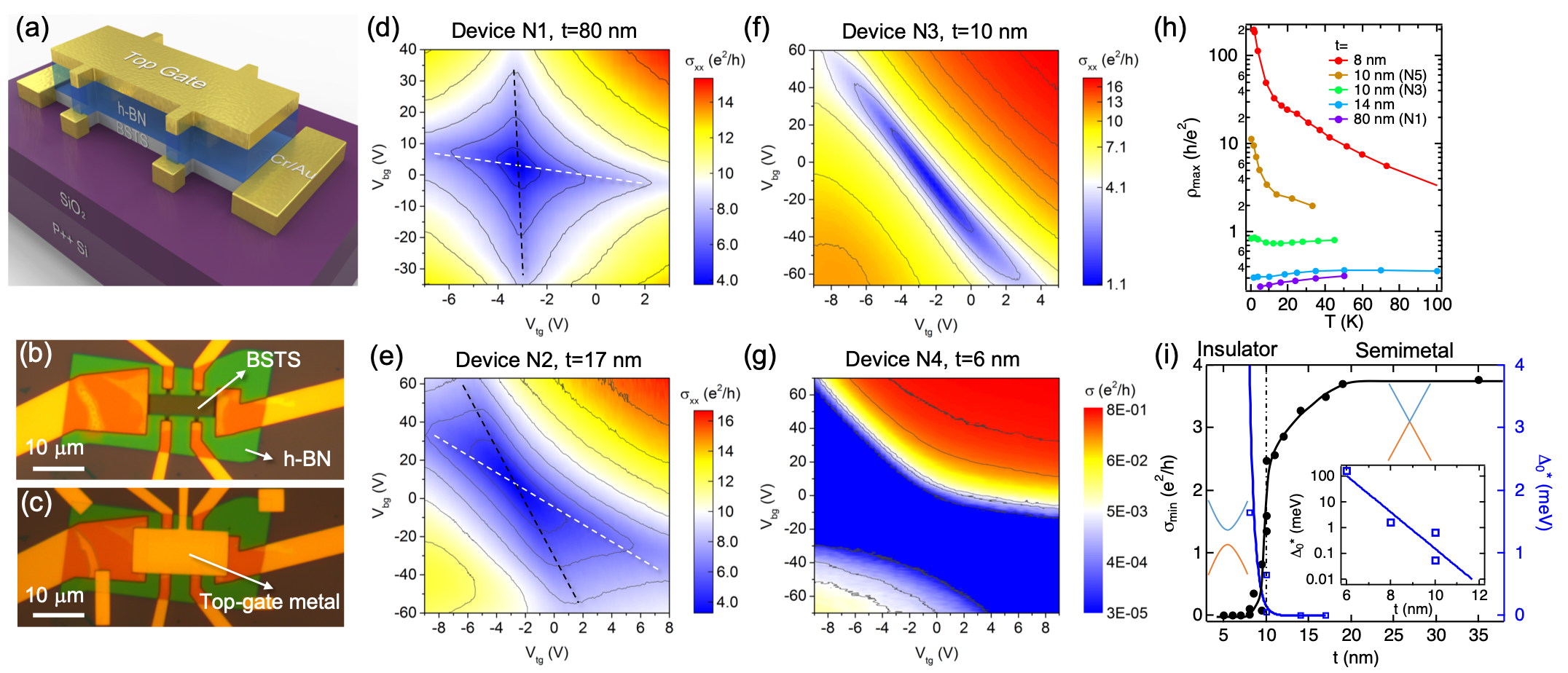}
	\vspace{-0.2in}
	\caption{(a) Schematic of a dual-gated TI device based on BSTS. (b,c) Optical images of Device N3 before (b) and after (c) top-gate metal deposition. (d-g) Measured conductivities (in color scale, with contours) as functions of $V_{tg}$ and $V_{bg}$ in four representative devices with decreasing thickness ($t$). In (d,e) the black/white dashed lines trace the top/bottom surface DPs. Data are measured at temperature $T=$ 0.35 K in (d-f), and at $T=$1.6 K in (g). (h) Temperature dependence of $\rho_{max}$ (log scale) for 5 representative devices. (i) The $\sigma_{min}$ ($=1/\rho_{max}$, left axis) at low $T$ ($<$2 K) and the extracted gap $\Delta_0^*$ (right axis) as functions of sample thickness $t$. The dashed-dotted vertical line marks the critical thickness $t_c=\sim$10 nm that separates the semimetal ($t>t_c$, corresponding to the 3D TI phase in the inset with gapless Dirac SS) and insulator ($t\leq t_c$, corresponding to the trivial insulator phase in the inset with gapped SS) behaviors. The inset plot shows $\Delta_0^*$ in log scale versus $t$ and an exponential fitting.}
	\vspace{-0.2in} 
\end{figure*}

Our experiment is based on a 3D TI crystal BSTS (BiSbTeSe$_2$) that has no detectable bulk conducting carriers at low temperature, with DPs of the topological SS exposed in the bulk band gap \cite{Xu2014a,Xu2016}, thus ideal for the study of low energy excitations in the vicinity of the surface DPs. The dual-gated BSTS devices \cite{Xu2016} were fabricated into Hall-bar structures (with channel length $l$, width $w$, thickness $t$) on highly p-doped Si substrates (with 300 nm-thick SiO$_2$ coating). Hexagonal boron nitride (h-BN) flakes (tens of nm in thickness) are transferred onto the devices as top-gate dielectrics (see a typical device schematic in Fig. 1a and optical images of Device N3 in Fig. 1b, c). Top and back gate voltages (denoted as $V_{tg}$  and $V_{bg}$) relative to the BSTS flake are applied to the top-gate metal and the doped Si, respectively. Upon dual-gating, the carrier types and densities of both the top and bottom surfaces, thus the measured conductivity, can be modulated. By reducing the thickness of the BSTS flake, the capacitive coupling between the two surfaces becomes stronger \cite{Kim2012c,Fatemi2014}. As it can be seen in the color map of 2D conductivity ($\sigma_{xx}=l/(wR_{xx}$), with $R_{xx}$ being the longitudinal resistance) versus $V_{tg}$ and $V_{bg}$ measured at low temperature, the black and white dashed lines tracing the DPs of top and bottom surfaces tend to merge together when the thickness t is reduced from 80 nm to 17 nm (Fig. 1d and 1e). Further reducing $t$ to $\sim$10 nm results in the DPs from the two surfaces to become indistinguishable (Fig. 1f). When the sample is only a few nm thick (e.g., Device N4 with $t=6$ nm in Fig. 1g), a hard gap opens, as indicated by the highly insulating (two-terminal conductivity $\sigma\ll e^2/h$) blue region.

The minimum conductivity $\sigma_{min}$ and maximum resistivity $\rho_{max}$ (=1/$\sigma_{min}$) are reached when the two surfaces are gated simultaneously to charge neutrality or DPs. In Fig. 1h, we plotted $\rho_{max}$ as a function of temperature ($T$) for a few representative samples. At $t>10$ nm, $\rho_{max}$ shows a metallic behavior ($d\rho_{max}/dT>0$), implying a zero or negligible gap. However, at $t<10$ nm, a strong insulating behavior ($d\rho_{max}/dT<0$) is observed. Around $t=10$ nm, different samples can behave differently. For example, while device N5 exhibits an insulating behavior, another device N3 exhibits a non-monotonic temperature dependence with its $\rho_{max}(T)$ close to $h/e^2$ and separating curves with metallic and insulating behaviors. It is consistent with the general observation from previous studies that the critical resistivity for metal-insulator transition in 2D electron systems is on the order of the resistance quantum $h/e^2$ \cite{DasSarma2014}. Fig. 1i shows $\sigma_{min}$ at base temperatures ($T\le 1.6$ K) for samples with various thicknesses. At large $t$ ($>\sim$20 nm), $\sigma_{min}$ saturates around a value close to $4e^2/h$ \cite{Xu2016}. The $\sigma_{min}$ starts to decrease below 20 nm and drops abruptly to zero below $\sim10$ nm. For samples that exhibits insulating behaviors, their $\rho_{max}(T)$ were fitted to thermal activation behavior $\rho_{max}(T)\propto e^{\Delta_0^*/2k_BT}$ (with $k_B$ being the Boltzmann constant) over appropriate temperature ranges to extract (see SI for details) the non-zero gap $\Delta_0^*$, plotted on the right axis of Fig. 1i. The $\Delta_0^*$ grows by about an order of magnitude when $t$ is reduced by $\sim$1.4 nm (see the exponential fitting in the inset of Fig. 1i), comparable to what was found for Bi$_2$Se$_3$ \cite{Zhang2010,Kim2013c}. Our data suggest that a measurable transport gap $\Delta_0^*$ (presumably driven by the inter-surface hybridization) opens at the DPs below a critical thickness $t_c=10\pm1$ nm in our samples.

\begin{figure}\label{fig:2}
	\centering\includegraphics[width=1\columnwidth]{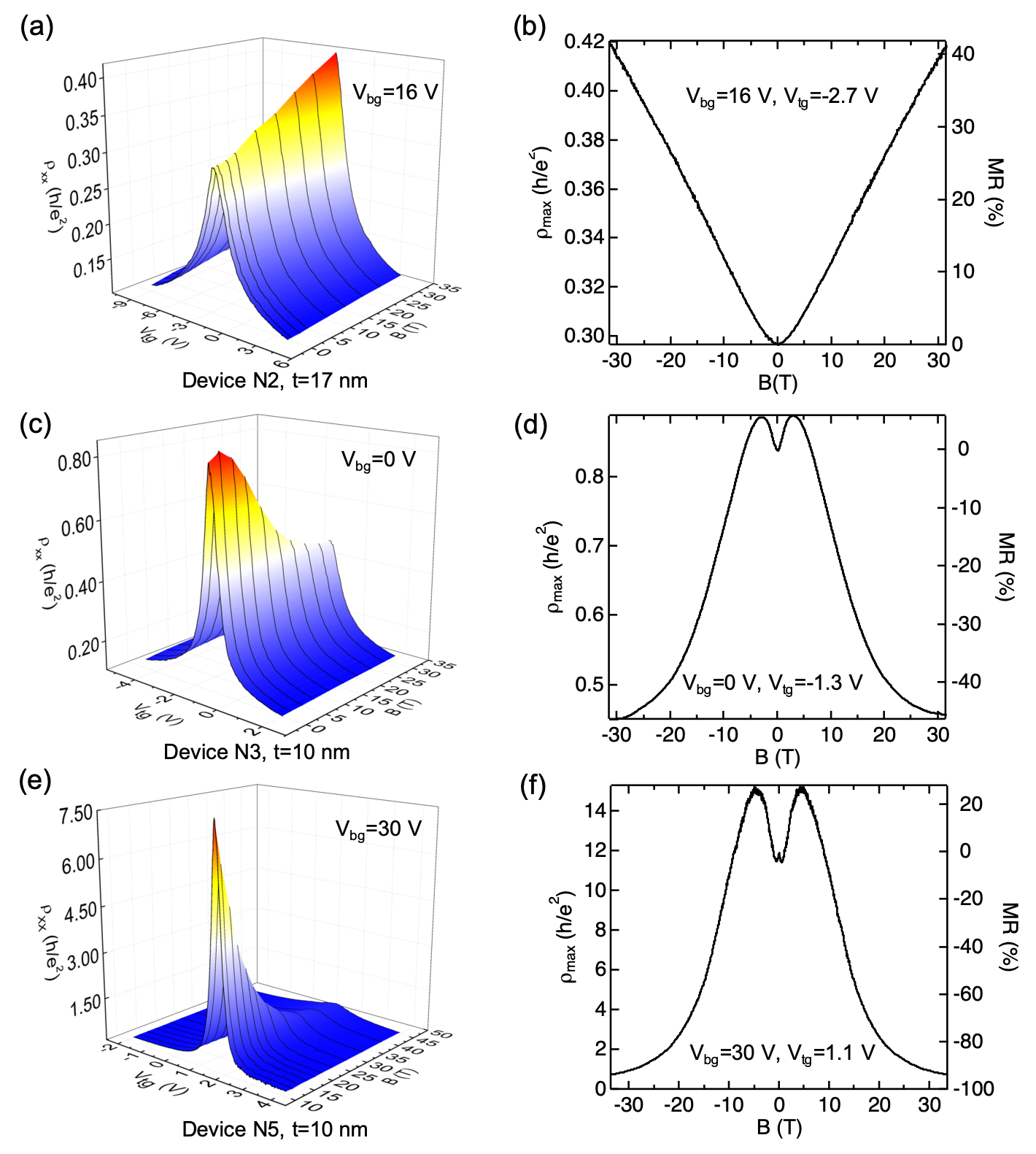}
	\vspace{-0.2in}
	\caption{(a,c,e) Resistivity of three representative devices (N2, N3 and N5) measured as a function of $V_{tg}$ at various in-plane $B$ fields at $T=$0.3 K. The corresponding $V_{bg}$'s are carefully tuned such that the $V_{tg}$ sweeps go through $\rho_{max}$. (b,d,f) The $\rho_{max}$ (and the corresponding MR, right axis) measured (at fixed $V_{bg}$ and $V_{tg}$ labeled in the figure) as a function of in-plane $B$ field for the three devices (N2, N3 and N5 respectively). }
	\vspace{-0.2in} 
\end{figure}

We have found that the resistances of the thicker and thinner samples respond to the in-plane magnetic field differently at low temperatures. For consistency, the samples are mounted with current direction parallel to $B$ (unless otherwise specified). We have measured multiple samples by either sweeping $V_{tg}$ (with $V_{bg}$ carefully tuned and then fixed at voltages such that these $V_{tg}$ sweeps go through $\rho_{max}$) at different in-plane $B$ fields, or measuring $\rho_{max}$ versus in-plane $B$ at fixed gate voltages. For relatively thick samples such as Device N2 with $t=17$ nm$>t_c$, the in-plane field up to $\sim$31 T only induced a relatively small positive MR of $\sim$40\% (Fig. 2a and 2b, noting $\rho_{max}(B)$ is approximately proportional to $B^2$ at low fields and to $B$ at higher fields). At low fields ($<\sim5$ T), thinner devices N3 and N5 (both $\sim$10 nm) also show some positive MR (for N5, we also observed an additional tiny cusp with negative MR near 0 T). Such low-field features in thinner devices disappear when we increase the temperature to just a few Kelvin (see SI), thus are attributed to phase coherent transport \cite{Lin2013,Liao2015}. In the following, we mainly focus on the higher field data showing a giant negative MR that has only been observed in thin samples with insulating behavior (hybridization gapped). For example, in Device N3 (Fig. 2c and 2d), $\rho_{max}$ drops dramatically above $\sim$5 T and saturates at high field ($\sim$30 T) to $\sim0.45h/e^2$. Notably for the more insulating sample N5 (Fig. 2e and 2f), $\rho_{max}$ drops by a factor of $\sim$20 (giving an MR$\sim$-95\%) from a very resistive value of $\sim12h/e^2$ at $B=0$ T to a value ($\sim0.55h/e^2$) again close to $h/2e^2$ at $B=45$ T. We have verified that Device N5 also showed a large negative MR (-85\% at 31 T) when the in-plane $B$ field is perpendicular to the current direction (SI Fig. S6). This contrasts with the negative MR associated with chiral anomaly in 3D Dirac/Weyl semimetals \cite{Xiong2015} and with various scattering mechanisms \cite{Goswami2015,Wiedmann2016}, as in those cases the negative MR disappears when the current is orthogonal to $B$. The field and temperature dependences we observed, as further discussed below, are also different from the behavior due to quantum interference effect in a variable-range-hopping regime \cite{Sivan1988}.

\begin{figure}\label{fig:3}
	\centering\includegraphics[width=1\columnwidth]{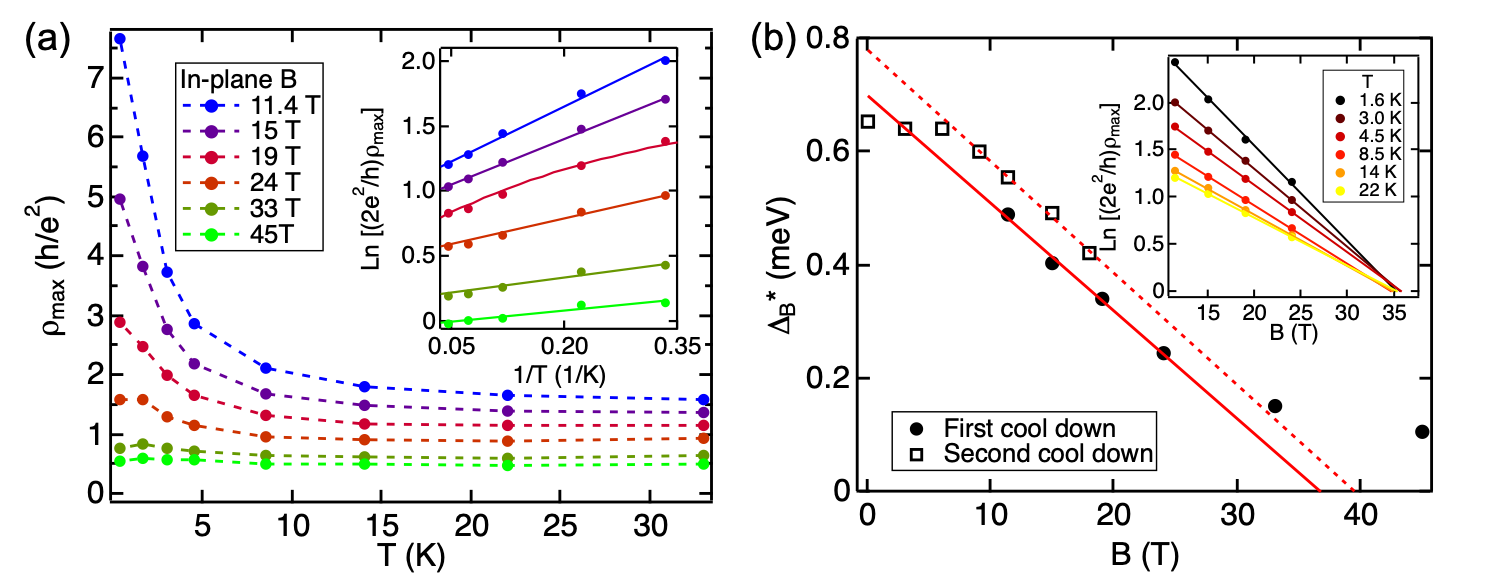}
	\vspace{-0.2in}
	\caption{(a) The $\rho_{max}$ of device N5 ($t\sim$10 nm) vs. $T$ at different in-plane B from 11.4 T up to 45 T. Inset shows corresponding thermal activation ($\rho_{max}(T)\propto e^{\Delta_B^*/2k_BT}$) fittings, while the extracted $\Delta_B^*$ is plotted (filled circles) as a function of $B$ in (b) along with data measured from another (second) cool down. (b) The linear fits for both cool downs indicate a gap closing at $B_c$ between 35 to 40 T, consistent with the inset showing the converging (at $B_c$=36 T) of all the linear fittings from Ln$[(2e^2/h)\rho_{max}]$ versus $B$ at different temperatures.}
	\vspace{-0.2in} 
\end{figure}

We performed systematic $V_{tg}$ sweeps (fixed $V_{bg}=30$ V) to extract $\rho_{max}$ with temperatures at various in-plane $B$ fields from 11.4 T to 45 T in Device N5. As shown in Fig. 3a, the insulating behavior of $\rho_{max}(T)$ is strongly suppressed at higher fields. At the highest field (45 T), $\rho_{max}$ saturates to a value close to $\sim h/2e^2$ and becomes relatively insensitive to temperature. We estimated the thermal activation gap $\Delta_B^*$ from the slope of Ln$[(2e^2/h)\rho_{max}$] versus $1/T$ in the temperature range of 3 K to 22 K (Fig. 3a inset) and plotted it versus the corresponding $B$ in Fig. 3b, which also displays the gap size measured in another (second) cool down for $B$ up to 18 T. The gap size $\Delta_B^*$ is found to differ slightly over different cool-downs but exhibits a similar dependence on $B$ in the intermediate field range (5 T$\sim$30 T).

Extrapolating the linear fits in Fig. 3b to zero suggests that the gap would close at a critical field ($B_c$) between 36 T to 40 T, around which we observe the sample (N5) to become metallic ($d\rho_{max}/dT>0$, see Fig. 3a) below $T\sim2$ K. However, some non-metallic behavior ($d\rho_{max}/dT<0$) can still be observed between 2 K to 22 K even at the highest fields (Fig. 3a) and fitted to a thermal activation, giving data points that deviate from the red solid line (Fig. 3b). A non-metallic behavior under large in-plane magnetic fields was also observed in gapless samples such as N2 with $t=17$ nm (SI Fig. S8b). The reason for this behavior remains to be understood. We have also verified that Ln$[(2e^2/h)\rho_{max}]$ of sample N5 is linear with $B$ ($<\sim25$ T) at different temperatures and all the fitted lines converge to a critical field of $\sim$36 T (inset of Fig. 3b). This also suggests $\Delta_B^*\propto(B_c-B)$, with a saturation resistivity $\sim h/2e^2$ (when $\Delta_B^*\sim0$) and gap closing at $B_c\sim$36 T at fixed temperatures.

\begin{figure}\label{fig:4}
	\centering\includegraphics[width=0.95\columnwidth]{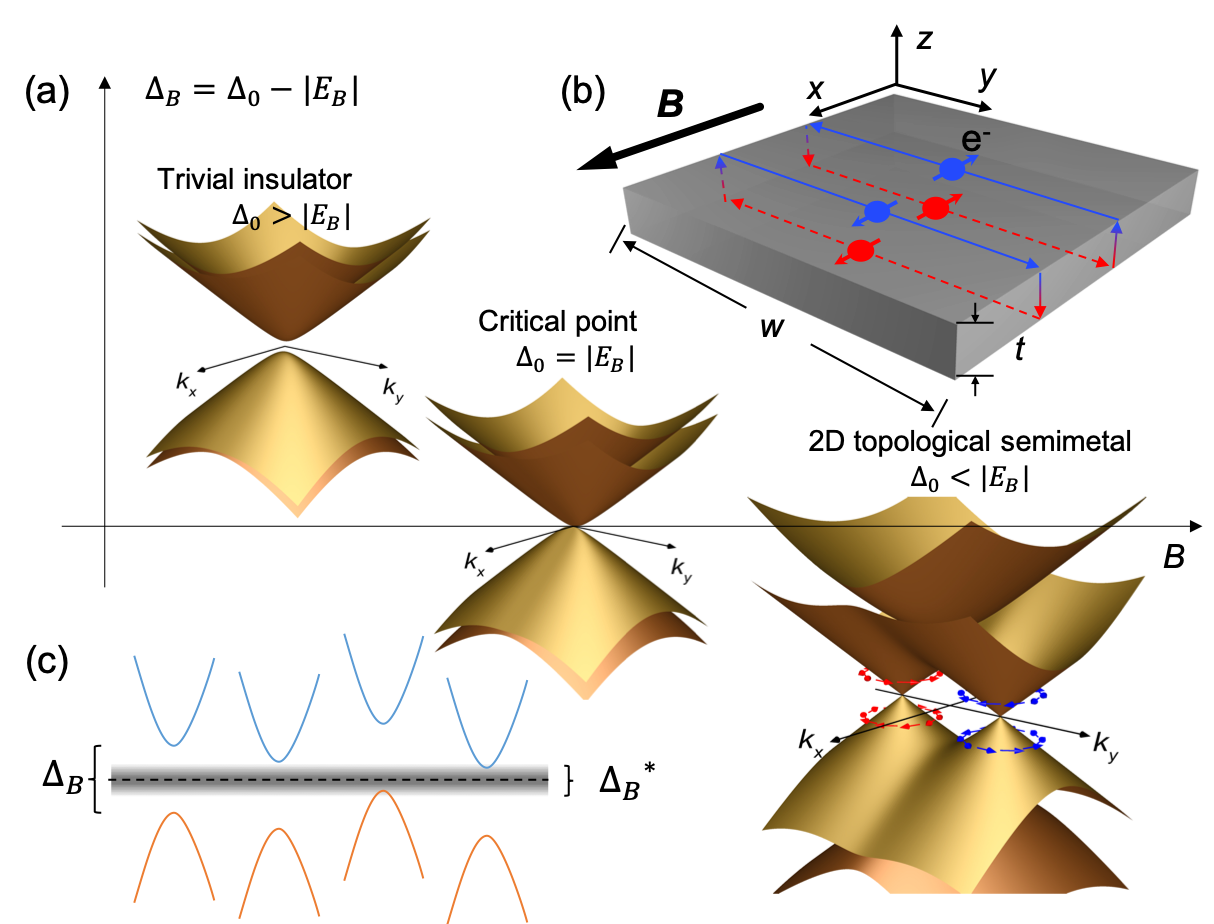}
	\vspace{-0.2in}
	\caption{(a) Predicted evolution of the surface band structure from a trivial insulator towards a 2D TSM, upon increasing in-plane $B$ field in a thin TI sample with hybridized SS. Red/blue arrows depict spin textures of the gapless Dirac cones, separated in the momentum space for 2D TSM, respectively. (b) Schematic of the orbital motion (normal to $B$ and $x$ direction) of spin-helical SS electrons for a 3D TI film. (c) The gap size extracted from thermal activation can be underestimated (resulting in a smaller measured value $\Delta_B^*=\Delta_B-\delta$) compared to the real gap $\Delta_B$ due to the smearing effect of disorder induced potential fluctuations ($\delta$) at different positions.}
	\vspace{-0.2in} 
\end{figure}

Such large negative MR and the corresponding insulator-semimetal transition have not been reported before in thin nonmagnetic 3D TIs with surface dominant conduction. Our observations of distinct behavior between thick and thin BSTS samples may be interpreted in terms of a theoretical prediction by A. A. Zyuzin et al. \cite{Zyuzin2011a}. Generally, in thick TIs the in-plane magnetic field $B$ (set to be along the $x$ direction) can introduce opposite shifts (along $k_y$) of top and bottom surface Dirac cones in the momentum space. This does not produce any MR effect in thick 3D TIs but will prevent the two DPs from annihilation and will tend to eliminate the hybridization gap in thin TIs (schematics shown in Fig. 4a). Semiclassically, a spin-helical SS electron with spin orientated along the $B$ field (thus spin magnetic moment $-g\mu_B/2$, with $g$ being the in-plane spin g-factor and $\mu_B$ the Bohr magneton) moves clockwise around the circumference (Fig. 4b) with orbital magnetic moment (parallel to $B$ field) $\mu_{orb}=\frac{e}{2(w+t)/v_f}wt\approx etv_f/2$ (noting width $w\gg t$ in our samples) \cite{Minot2004}. Both the spin and orbital magnetic moments couple to the $B$ field, giving rise to an effective Zeeman energy $E_B=g_{eff} \mu_B B=(g\mu_B-etv_F)B$ (total effective $g_{eff}=g-etv_F/\mu_B$, the second term can also be considered as due to the Aharonov-Bohm phase gradient between the two opposite surfaces). In thin TIs with hybridization gap $\Delta_0$ (at zero $B$ field), one can show (Ref. \cite{Zyuzin2011a} and SI) that the massive Dirac band is spin-split by the above ``effective Zeeman energy'', shrinking the hybridization gap linearly with $B$ as $\Delta_B=\Delta_0-|E_B|$. The gap vanishes at a critical field $B_c$ ($|E_B|=\Delta_0$), where the dispersion near $k=0$ becomes quadratic along $k_y$ and linear along $k_x$ (see Fig. 4a). With further increasing $B$ ($>B_c$), two DPs are restored and separated by $\sqrt{E_B^2-\Delta_0^2}/\hbar v_f$ along $k_y$. This gives rise to a distinct 2D TSM that is topologically stable as long as translational symmetry is preserved \cite{Zyuzin2011a}.

The above mechanism can qualitatively explain the trend we observed in experiments. However, the slope of the linear fitting yields the gap closing rate $|\frac{E_B}{B}|\approx0.02$ meV/T (corresponding to $g_{eff}\approx0.33$). This is nearly two orders of magnitudes smaller compared with $|\frac{E_B}{B}|\approx1$ meV/T theoretically estimated for a 10-nm sample by A. A. Zyuzin et al. \cite{Zyuzin2011a}, which assumed $g=2$, leading to a negligibly small contribution from the Zeeman effect. Our results imply a large $g$, giving rise to a Zeeman term ($g\mu_BB$) that is comparable with the orbital term ($etv_fB$). Thus, the two nearly cancel to give a small $|\frac{E_B}{B}|$. Assuming a typical $v_f=3.5\times10^5$ m/s for topological SS with purely linear dispersion, we get a in-plane SS g-factor of $\sim60$. In actual 3D TI materials such as BSTS, the surface Dirac cone contains substantial nonlinearity that can be described by a quadratic mass term added to the SS Hamiltonian. Subsequently, a reduced $v_f\approx1.3\times10^5$ m/s, which describes the linear part in the Hamiltonian, yields a g-factor of $\sim20$ (SI). It has been pointed out that the Zeeman coupling of the SS carriers can be highly anisotropic \cite{Chu2011}. In previous experiments, only an out-of-plane SS g-factor is determined and found to vary significantly in different TI materials \cite{Taskin2011a,Fu2016}. Our study provides a method to extract the \textit{in-plane} g-factor of SS carriers. 

We have to note that in our experiments, the gap extracted from thermal activation is an effective transport gap ($\Delta_B^*$) and can be smaller than the real band gap ($\Delta_B$) due to disorder-induced smearing, namely $\Delta_B^*=\Delta_B-\delta$, where $\delta$ is a correction due to the potential fluctuations (likely to be on the order of several meV or higher \cite{Nam2017}) in the system (Fig. 4c). Therefore, the observed apparent metallic behavior ($\Delta_B^*$ reaching 0) in Device N5 above $B_c\sim$36 T does not necessarily indicate the realization of the 2D TSM phase, which requires closing the real gap $\Delta_B$  and possibly much larger magnetic field than $B_c$ (noting the relatively small gap-closing rate of 0.02 meV/T in light of the estimated $\delta\sim$meV in our BSTS samples).  It might be easier to realize the 2D TSM phase (at lower $B$ field) in other TI systems with a smaller or even negative g-factor (so the gap closing rate can be much larger than that in our samples). It would also be interesting for future studies to clarify whether the saturation resistivity $\sim h/2e^2$ is related to the modification of band structure and magnetic field induced spin-flip scatterings \cite{Taskin2017}.

To summarize, we have demonstrated in ultrathin BSTS films with hybridized and gapped surface states a transition from an insulator to semimetal induced by either increasing thickness or an in-plane magnetic field. The in-plane magnetic field can shrink the hybridization gap and give a large negative MR that may be exploited for applications. Sufficient in-plane magnetic field is predicted to drive the thin 3D TI with hybridization gap to a 2D TSM phase, which would have 2 single-fold Dirac cones separated in the momentum space and provide a 2D analogue of Weyl semimetal (even though Weyl fermions cannot be strictly defined in even spatial dimensions \cite{Armitage2018}). Such a TSM (with the momentum space displacement between Dirac cones tunable by the in-plane magnetic field) can possess interesting 1D edge states \cite{CastroNeto2009,Takahashi2015}, which are analogous to the Fermi arcs in 3D Weyl semimetals \cite{Armitage2018} and have signatures that future experiments (e.g. performed at even higher magnetic fields) can search as evidence for the TSMs. 

\begin{acknowledgments}
We thank A. Suslov, T. Murphy, J-H. Park and Z. Lu for experimental assistance, and Y. Jiang, C. Liu and K. T. Law for discussions. This work has benefited from partial support from DARPA MESO program (Grant N66001-11-1-4107) and NSF (Grant DMR \#1410942 and EFMA \#1641101). G. Jiang and R. Biswas were supported by Purdue University startup funds. A portion of this work was performed at the National High Magnetic Field Laboratory, which is supported by NSF Cooperative Agreement No. DMR-1157490, the State of Florida, and the U.S. Department of Energy.
\end{acknowledgments}

\bibliographystyle{apsrev4-1}
\bibliography{bibliography}

\begin{thebibliography}{35}%
\makeatletter
\providecommand \@ifxundefined [1]{%
 \@ifx{#1\undefined}
}%
\providecommand \@ifnum [1]{%
 \ifnum #1\expandafter \@firstoftwo
 \else \expandafter \@secondoftwo
 \fi
}%
\providecommand \@ifx [1]{%
 \ifx #1\expandafter \@firstoftwo
 \else \expandafter \@secondoftwo
 \fi
}%
\providecommand \natexlab [1]{#1}%
\providecommand \enquote  [1]{``#1''}%
\providecommand \bibnamefont  [1]{#1}%
\providecommand \bibfnamefont [1]{#1}%
\providecommand \citenamefont [1]{#1}%
\providecommand \href@noop [0]{\@secondoftwo}%
\providecommand \href [0]{\begingroup \@sanitize@url \@href}%
\providecommand \@href[1]{\@@startlink{#1}\@@href}%
\providecommand \@@href[1]{\endgroup#1\@@endlink}%
\providecommand \@sanitize@url [0]{\catcode `\\12\catcode `\$12\catcode
  `\&12\catcode `\#12\catcode `\^12\catcode `\_12\catcode `\%12\relax}%
\providecommand \@@startlink[1]{}%
\providecommand \@@endlink[0]{}%
\providecommand \url  [0]{\begingroup\@sanitize@url \@url }%
\providecommand \@url [1]{\endgroup\@href {#1}{\urlprefix }}%
\providecommand \urlprefix  [0]{URL }%
\providecommand \Eprint [0]{\href }%
\providecommand \doibase [0]{http://dx.doi.org/}%
\providecommand \selectlanguage [0]{\@gobble}%
\providecommand \bibinfo  [0]{\@secondoftwo}%
\providecommand \bibfield  [0]{\@secondoftwo}%
\providecommand \translation [1]{[#1]}%
\providecommand \BibitemOpen [0]{}%
\providecommand \bibitemStop [0]{}%
\providecommand \bibitemNoStop [0]{.\EOS\space}%
\providecommand \EOS [0]{\spacefactor3000\relax}%
\providecommand \BibitemShut  [1]{\csname bibitem#1\endcsname}%
\let\auto@bib@innerbib\@empty
\bibitem [{\citenamefont {{Castro Neto}}\ \emph {et~al.}(2009)\citenamefont
  {{Castro Neto}}, \citenamefont {Guinea}, \citenamefont {Peres}, \citenamefont
  {Novoselov},\ and\ \citenamefont {Geim}}]{CastroNeto2009}%
  \BibitemOpen
  \bibfield  {author} {\bibinfo {author} {\bibfnamefont {A.~H.}\ \bibnamefont
  {{Castro Neto}}}, \bibinfo {author} {\bibfnamefont {F.}~\bibnamefont
  {Guinea}}, \bibinfo {author} {\bibfnamefont {N.~M.~R.}\ \bibnamefont
  {Peres}}, \bibinfo {author} {\bibfnamefont {K.~S.}\ \bibnamefont
  {Novoselov}}, \ and\ \bibinfo {author} {\bibfnamefont {A.~K.}\ \bibnamefont
  {Geim}},\ }\href {\doibase 10.1103/RevModPhys.81.109} {\bibfield  {journal}
  {\bibinfo  {journal} {Reviews of Modern Physics}\ }\textbf {\bibinfo {volume}
  {81}},\ \bibinfo {pages} {109} (\bibinfo {year} {2009})}\BibitemShut
  {NoStop}%
\bibitem [{\citenamefont {Hasan}\ and\ \citenamefont {Kane}(2010)}]{Hasan2010}%
  \BibitemOpen
  \bibfield  {author} {\bibinfo {author} {\bibfnamefont {M.~Z.}\ \bibnamefont
  {Hasan}}\ and\ \bibinfo {author} {\bibfnamefont {C.~L.}\ \bibnamefont
  {Kane}},\ }\href {\doibase 10.1103/RevModPhys.82.3045} {\bibfield  {journal}
  {\bibinfo  {journal} {Reviews of Modern Physics}\ }\textbf {\bibinfo {volume}
  {82}},\ \bibinfo {pages} {3045} (\bibinfo {year} {2010})}\BibitemShut
  {NoStop}%
\bibitem [{\citenamefont {Qi}\ and\ \citenamefont {Zhang}(2011)}]{Qi2011}%
  \BibitemOpen
  \bibfield  {author} {\bibinfo {author} {\bibfnamefont {X.-L.}\ \bibnamefont
  {Qi}}\ and\ \bibinfo {author} {\bibfnamefont {S.-C.}\ \bibnamefont {Zhang}},\
  }\href {\doibase 10.1103/RevModPhys.83.1057} {\bibfield  {journal} {\bibinfo
  {journal} {Reviews of Modern Physics}\ }\textbf {\bibinfo {volume} {83}},\
  \bibinfo {pages} {1057} (\bibinfo {year} {2011})}\BibitemShut {NoStop}%
\bibitem [{\citenamefont {Armitage}\ \emph {et~al.}(2018)\citenamefont
  {Armitage}, \citenamefont {Mele},\ and\ \citenamefont
  {Vishwanath}}]{Armitage2018}%
  \BibitemOpen
  \bibfield  {author} {\bibinfo {author} {\bibfnamefont {N.~P.}\ \bibnamefont
  {Armitage}}, \bibinfo {author} {\bibfnamefont {E.~J.}\ \bibnamefont {Mele}},
  \ and\ \bibinfo {author} {\bibfnamefont {A.}~\bibnamefont {Vishwanath}},\
  }\href {\doibase 10.1103/RevModPhys.90.015001} {\bibfield  {journal}
  {\bibinfo  {journal} {Reviews of Modern Physics}\ }\textbf {\bibinfo {volume}
  {90}},\ \bibinfo {pages} {015001} (\bibinfo {year} {2018})}\BibitemShut
  {NoStop}%
\bibitem [{\citenamefont {Lu}\ \emph {et~al.}(2014)\citenamefont {Lu},
  \citenamefont {Joannopoulos},\ and\ \citenamefont
  {Solja{\v{c}}i{\'{c}}}}]{Lu2014}%
  \BibitemOpen
  \bibfield  {author} {\bibinfo {author} {\bibfnamefont {L.}~\bibnamefont
  {Lu}}, \bibinfo {author} {\bibfnamefont {J.~D.}\ \bibnamefont
  {Joannopoulos}}, \ and\ \bibinfo {author} {\bibfnamefont {M.}~\bibnamefont
  {Solja{\v{c}}i{\'{c}}}},\ }\href {\doibase 10.1038/nphoton.2014.248}
  {\bibfield  {journal} {\bibinfo  {journal} {Nature Photonics}\ }\textbf
  {\bibinfo {volume} {8}},\ \bibinfo {pages} {821} (\bibinfo {year}
  {2014})}\BibitemShut {NoStop}%
\bibitem [{\citenamefont {Bellec}\ \emph {et~al.}(2013)\citenamefont {Bellec},
  \citenamefont {Kuhl}, \citenamefont {Montambaux},\ and\ \citenamefont
  {Mortessagne}}]{Bellec2013}%
  \BibitemOpen
  \bibfield  {author} {\bibinfo {author} {\bibfnamefont {M.}~\bibnamefont
  {Bellec}}, \bibinfo {author} {\bibfnamefont {U.}~\bibnamefont {Kuhl}},
  \bibinfo {author} {\bibfnamefont {G.}~\bibnamefont {Montambaux}}, \ and\
  \bibinfo {author} {\bibfnamefont {F.}~\bibnamefont {Mortessagne}},\ }\href
  {\doibase 10.1103/PhysRevLett.110.033902} {\bibfield  {journal} {\bibinfo
  {journal} {Physical Review Letters}\ }\textbf {\bibinfo {volume} {110}},\
  \bibinfo {pages} {033902} (\bibinfo {year} {2013})}\BibitemShut {NoStop}%
\bibitem [{\citenamefont {Tarruell}\ \emph {et~al.}(2012)\citenamefont
  {Tarruell}, \citenamefont {Greif}, \citenamefont {Uehlinger}, \citenamefont
  {Jotzu},\ and\ \citenamefont {Esslinger}}]{Tarruell2012}%
  \BibitemOpen
  \bibfield  {author} {\bibinfo {author} {\bibfnamefont {L.}~\bibnamefont
  {Tarruell}}, \bibinfo {author} {\bibfnamefont {D.}~\bibnamefont {Greif}},
  \bibinfo {author} {\bibfnamefont {T.}~\bibnamefont {Uehlinger}}, \bibinfo
  {author} {\bibfnamefont {G.}~\bibnamefont {Jotzu}}, \ and\ \bibinfo {author}
  {\bibfnamefont {T.}~\bibnamefont {Esslinger}},\ }\href {\doibase
  10.1038/nature10871} {\bibfield  {journal} {\bibinfo  {journal} {Nature}\
  }\textbf {\bibinfo {volume} {483}},\ \bibinfo {pages} {302} (\bibinfo {year}
  {2012})}\BibitemShut {NoStop}%
\bibitem [{\citenamefont {Wehling}\ \emph {et~al.}(2014)\citenamefont
  {Wehling}, \citenamefont {Black-Schaffer},\ and\ \citenamefont
  {Balatsky}}]{Wehling2014}%
  \BibitemOpen
  \bibfield  {author} {\bibinfo {author} {\bibfnamefont {T.}~\bibnamefont
  {Wehling}}, \bibinfo {author} {\bibfnamefont {A.}~\bibnamefont
  {Black-Schaffer}}, \ and\ \bibinfo {author} {\bibfnamefont {A.}~\bibnamefont
  {Balatsky}},\ }\href {\doibase 10.1080/00018732.2014.927109} {\bibfield
  {journal} {\bibinfo  {journal} {Advances in Physics}\ }\textbf {\bibinfo
  {volume} {63}},\ \bibinfo {pages} {1} (\bibinfo {year} {2014})}\BibitemShut
  {NoStop}%
\bibitem [{\citenamefont {Hasegawa}\ \emph {et~al.}(2006)\citenamefont
  {Hasegawa}, \citenamefont {Konno}, \citenamefont {Nakano},\ and\
  \citenamefont {Kohmoto}}]{Hasegawa2006}%
  \BibitemOpen
  \bibfield  {author} {\bibinfo {author} {\bibfnamefont {Y.}~\bibnamefont
  {Hasegawa}}, \bibinfo {author} {\bibfnamefont {R.}~\bibnamefont {Konno}},
  \bibinfo {author} {\bibfnamefont {H.}~\bibnamefont {Nakano}}, \ and\ \bibinfo
  {author} {\bibfnamefont {M.}~\bibnamefont {Kohmoto}},\ }\href {\doibase
  10.1103/PhysRevB.74.033413} {\bibfield  {journal} {\bibinfo  {journal}
  {Physical Review B}\ }\textbf {\bibinfo {volume} {74}},\ \bibinfo {pages}
  {033413} (\bibinfo {year} {2006})}\BibitemShut {NoStop}%
\bibitem [{\citenamefont {Baik}\ \emph {et~al.}(2015)\citenamefont {Baik},
  \citenamefont {Kim}, \citenamefont {Yi},\ and\ \citenamefont
  {Choi}}]{Baik2015}%
  \BibitemOpen
  \bibfield  {author} {\bibinfo {author} {\bibfnamefont {S.~S.}\ \bibnamefont
  {Baik}}, \bibinfo {author} {\bibfnamefont {K.~S.}\ \bibnamefont {Kim}},
  \bibinfo {author} {\bibfnamefont {Y.}~\bibnamefont {Yi}}, \ and\ \bibinfo
  {author} {\bibfnamefont {H.~J.}\ \bibnamefont {Choi}},\ }\href {\doibase
  10.1021/acs.nanolett.5b04106} {\bibfield  {journal} {\bibinfo  {journal}
  {Nano Letters}\ }\textbf {\bibinfo {volume} {15}},\ \bibinfo {pages} {7788}
  (\bibinfo {year} {2015})}\BibitemShut {NoStop}%
\bibitem [{\citenamefont {Pereira}\ \emph {et~al.}(2009)\citenamefont
  {Pereira}, \citenamefont {{Castro Neto}},\ and\ \citenamefont
  {Peres}}]{Pereira2009}%
  \BibitemOpen
  \bibfield  {author} {\bibinfo {author} {\bibfnamefont {V.~M.}\ \bibnamefont
  {Pereira}}, \bibinfo {author} {\bibfnamefont {A.~H.}\ \bibnamefont {{Castro
  Neto}}}, \ and\ \bibinfo {author} {\bibfnamefont {N.~M.~R.}\ \bibnamefont
  {Peres}},\ }\href {\doibase 10.1103/PhysRevB.80.045401} {\bibfield  {journal}
  {\bibinfo  {journal} {Physical Review B}\ }\textbf {\bibinfo {volume} {80}},\
  \bibinfo {pages} {045401} (\bibinfo {year} {2009})}\BibitemShut {NoStop}%
\bibitem [{\citenamefont {Feilhauer}\ \emph {et~al.}(2015)\citenamefont
  {Feilhauer}, \citenamefont {Apel},\ and\ \citenamefont
  {Schweitzer}}]{Feilhauer2015}%
  \BibitemOpen
  \bibfield  {author} {\bibinfo {author} {\bibfnamefont {J.}~\bibnamefont
  {Feilhauer}}, \bibinfo {author} {\bibfnamefont {W.}~\bibnamefont {Apel}}, \
  and\ \bibinfo {author} {\bibfnamefont {L.}~\bibnamefont {Schweitzer}},\
  }\href {\doibase 10.1103/PhysRevB.92.245424} {\bibfield  {journal} {\bibinfo
  {journal} {Physical Review B}\ }\textbf {\bibinfo {volume} {92}},\ \bibinfo
  {pages} {245424} (\bibinfo {year} {2015})}\BibitemShut {NoStop}%
\bibitem [{\citenamefont {Kim}\ \emph {et~al.}(2015)\citenamefont {Kim},
  \citenamefont {Baik}, \citenamefont {Ryu}, \citenamefont {Sohn},
  \citenamefont {Park}, \citenamefont {Park}, \citenamefont {Denlinger},
  \citenamefont {Yi}, \citenamefont {Choi},\ and\ \citenamefont
  {Kim}}]{Kim2015}%
  \BibitemOpen
  \bibfield  {author} {\bibinfo {author} {\bibfnamefont {J.}~\bibnamefont
  {Kim}}, \bibinfo {author} {\bibfnamefont {S.~S.}\ \bibnamefont {Baik}},
  \bibinfo {author} {\bibfnamefont {S.~H.}\ \bibnamefont {Ryu}}, \bibinfo
  {author} {\bibfnamefont {Y.}~\bibnamefont {Sohn}}, \bibinfo {author}
  {\bibfnamefont {S.}~\bibnamefont {Park}}, \bibinfo {author} {\bibfnamefont
  {B.-G.}\ \bibnamefont {Park}}, \bibinfo {author} {\bibfnamefont
  {J.}~\bibnamefont {Denlinger}}, \bibinfo {author} {\bibfnamefont
  {Y.}~\bibnamefont {Yi}}, \bibinfo {author} {\bibfnamefont {H.~J.}\
  \bibnamefont {Choi}}, \ and\ \bibinfo {author} {\bibfnamefont {K.~S.}\
  \bibnamefont {Kim}},\ }\href {\doibase 10.1126/science.aaa6486} {\bibfield
  {journal} {\bibinfo  {journal} {Science}\ }\textbf {\bibinfo {volume}
  {349}},\ \bibinfo {pages} {723} (\bibinfo {year} {2015})}\BibitemShut
  {NoStop}%
\bibitem [{\citenamefont {Zhang}\ \emph {et~al.}(2010)\citenamefont {Zhang},
  \citenamefont {He}, \citenamefont {Chang}, \citenamefont {Song},
  \citenamefont {Wang}, \citenamefont {Chen}, \citenamefont {Jia},
  \citenamefont {Fang}, \citenamefont {Dai}, \citenamefont {Shan},
  \citenamefont {Shen}, \citenamefont {Niu}, \citenamefont {Qi}, \citenamefont
  {Zhang}, \citenamefont {Ma},\ and\ \citenamefont {Xue}}]{Zhang2010}%
  \BibitemOpen
  \bibfield  {author} {\bibinfo {author} {\bibfnamefont {Y.}~\bibnamefont
  {Zhang}}, \bibinfo {author} {\bibfnamefont {K.}~\bibnamefont {He}}, \bibinfo
  {author} {\bibfnamefont {C.-Z.}\ \bibnamefont {Chang}}, \bibinfo {author}
  {\bibfnamefont {C.-L.}\ \bibnamefont {Song}}, \bibinfo {author}
  {\bibfnamefont {L.-L.}\ \bibnamefont {Wang}}, \bibinfo {author}
  {\bibfnamefont {X.}~\bibnamefont {Chen}}, \bibinfo {author} {\bibfnamefont
  {J.-F.}\ \bibnamefont {Jia}}, \bibinfo {author} {\bibfnamefont
  {Z.}~\bibnamefont {Fang}}, \bibinfo {author} {\bibfnamefont {X.}~\bibnamefont
  {Dai}}, \bibinfo {author} {\bibfnamefont {W.-Y.}\ \bibnamefont {Shan}},
  \bibinfo {author} {\bibfnamefont {S.-Q.}\ \bibnamefont {Shen}}, \bibinfo
  {author} {\bibfnamefont {Q.}~\bibnamefont {Niu}}, \bibinfo {author}
  {\bibfnamefont {X.-L.}\ \bibnamefont {Qi}}, \bibinfo {author} {\bibfnamefont
  {S.-C.}\ \bibnamefont {Zhang}}, \bibinfo {author} {\bibfnamefont {X.-C.}\
  \bibnamefont {Ma}}, \ and\ \bibinfo {author} {\bibfnamefont {Q.-K.}\
  \bibnamefont {Xue}},\ }\href {http://dx.doi.org/10.1038/nphys1689} {\bibfield
   {journal} {\bibinfo  {journal} {Nature Physics}\ }\textbf {\bibinfo {volume}
  {6}},\ \bibinfo {pages} {584} (\bibinfo {year} {2010})}\BibitemShut {NoStop}%
\bibitem [{\citenamefont {Zyuzin}\ \emph {et~al.}(2011)\citenamefont {Zyuzin},
  \citenamefont {Hook},\ and\ \citenamefont {Burkov}}]{Zyuzin2011a}%
  \BibitemOpen
  \bibfield  {author} {\bibinfo {author} {\bibfnamefont {A.~A.}\ \bibnamefont
  {Zyuzin}}, \bibinfo {author} {\bibfnamefont {M.~D.}\ \bibnamefont {Hook}}, \
  and\ \bibinfo {author} {\bibfnamefont {A.~A.}\ \bibnamefont {Burkov}},\
  }\href {\doibase 10.1103/PhysRevB.83.245428} {\bibfield  {journal} {\bibinfo
  {journal} {Physical Review B}\ }\textbf {\bibinfo {volume} {83}},\ \bibinfo
  {pages} {245428} (\bibinfo {year} {2011})}\BibitemShut {NoStop}%
\bibitem [{\citenamefont {Wiedmann}\ \emph {et~al.}(2016)\citenamefont
  {Wiedmann}, \citenamefont {Jost}, \citenamefont {Fauqu{\'{e}}}, \citenamefont
  {{van Dijk}}, \citenamefont {Meijer}, \citenamefont {Khouri}, \citenamefont
  {Pezzini}, \citenamefont {Grauer}, \citenamefont {Schreyeck}, \citenamefont
  {Br{\"{u}}ne}, \citenamefont {Buhmann}, \citenamefont {Molenkamp},\ and\
  \citenamefont {Hussey}}]{Wiedmann2016}%
  \BibitemOpen
  \bibfield  {author} {\bibinfo {author} {\bibfnamefont {S.}~\bibnamefont
  {Wiedmann}}, \bibinfo {author} {\bibfnamefont {A.}~\bibnamefont {Jost}},
  \bibinfo {author} {\bibfnamefont {B.}~\bibnamefont {Fauqu{\'{e}}}}, \bibinfo
  {author} {\bibfnamefont {J.}~\bibnamefont {{van Dijk}}}, \bibinfo {author}
  {\bibfnamefont {M.~J.}\ \bibnamefont {Meijer}}, \bibinfo {author}
  {\bibfnamefont {T.}~\bibnamefont {Khouri}}, \bibinfo {author} {\bibfnamefont
  {S.}~\bibnamefont {Pezzini}}, \bibinfo {author} {\bibfnamefont
  {S.}~\bibnamefont {Grauer}}, \bibinfo {author} {\bibfnamefont
  {S.}~\bibnamefont {Schreyeck}}, \bibinfo {author} {\bibfnamefont
  {C.}~\bibnamefont {Br{\"{u}}ne}}, \bibinfo {author} {\bibfnamefont
  {H.}~\bibnamefont {Buhmann}}, \bibinfo {author} {\bibfnamefont {L.~W.}\
  \bibnamefont {Molenkamp}}, \ and\ \bibinfo {author} {\bibfnamefont {N.~E.}\
  \bibnamefont {Hussey}},\ }\href {\doibase 10.1103/PhysRevB.94.081302}
  {\bibfield  {journal} {\bibinfo  {journal} {Physical Review B}\ }\textbf
  {\bibinfo {volume} {94}},\ \bibinfo {pages} {081302(R)} (\bibinfo {year}
  {2016})}\BibitemShut {NoStop}%
\bibitem [{\citenamefont {Breunig}\ \emph {et~al.}(2017)\citenamefont
  {Breunig}, \citenamefont {Wang}, \citenamefont {Taskin}, \citenamefont {Lux},
  \citenamefont {Rosch},\ and\ \citenamefont {Ando}}]{Breunig2017}%
  \BibitemOpen
  \bibfield  {author} {\bibinfo {author} {\bibfnamefont {O.}~\bibnamefont
  {Breunig}}, \bibinfo {author} {\bibfnamefont {Z.}~\bibnamefont {Wang}},
  \bibinfo {author} {\bibfnamefont {A.~A.}\ \bibnamefont {Taskin}}, \bibinfo
  {author} {\bibfnamefont {J.}~\bibnamefont {Lux}}, \bibinfo {author}
  {\bibfnamefont {A.}~\bibnamefont {Rosch}}, \ and\ \bibinfo {author}
  {\bibfnamefont {Y.}~\bibnamefont {Ando}},\ }\href {\doibase
  10.1038/ncomms15545} {\bibfield  {journal} {\bibinfo  {journal} {Nature
  Communications}\ }\textbf {\bibinfo {volume} {8}},\ \bibinfo {pages} {15545}
  (\bibinfo {year} {2017})},\ \Eprint {http://arxiv.org/abs/1703.10784}
  {1703.10784} \BibitemShut {NoStop}%
\bibitem [{\citenamefont {Lin}\ \emph {et~al.}(2013)\citenamefont {Lin},
  \citenamefont {He}, \citenamefont {Liao}, \citenamefont {Wang}, \citenamefont
  {Sacksteder~IV}, \citenamefont {Yang}, \citenamefont {Guan}, \citenamefont
  {Zhang}, \citenamefont {Gu}, \citenamefont {Zhang}, \citenamefont {Zeng},
  \citenamefont {Dai}, \citenamefont {Wu},\ and\ \citenamefont {Li}}]{Lin2013}%
  \BibitemOpen
  \bibfield  {author} {\bibinfo {author} {\bibfnamefont {C.~J.}\ \bibnamefont
  {Lin}}, \bibinfo {author} {\bibfnamefont {X.~Y.}\ \bibnamefont {He}},
  \bibinfo {author} {\bibfnamefont {J.}~\bibnamefont {Liao}}, \bibinfo {author}
  {\bibfnamefont {X.~X.}\ \bibnamefont {Wang}}, \bibinfo {author}
  {\bibfnamefont {V.}~\bibnamefont {Sacksteder~IV}}, \bibinfo {author}
  {\bibfnamefont {W.~M.}\ \bibnamefont {Yang}}, \bibinfo {author}
  {\bibfnamefont {T.}~\bibnamefont {Guan}}, \bibinfo {author} {\bibfnamefont
  {Q.~M.}\ \bibnamefont {Zhang}}, \bibinfo {author} {\bibfnamefont
  {L.}~\bibnamefont {Gu}}, \bibinfo {author} {\bibfnamefont {G.~Y.}\
  \bibnamefont {Zhang}}, \bibinfo {author} {\bibfnamefont {C.~G.}\ \bibnamefont
  {Zeng}}, \bibinfo {author} {\bibfnamefont {X.}~\bibnamefont {Dai}}, \bibinfo
  {author} {\bibfnamefont {K.~H.}\ \bibnamefont {Wu}}, \ and\ \bibinfo {author}
  {\bibfnamefont {Y.~Q.}\ \bibnamefont {Li}},\ }\href {\doibase
  10.1103/PhysRevB.88.041307} {\bibfield  {journal} {\bibinfo  {journal}
  {Physical Review B}\ }\textbf {\bibinfo {volume} {88}},\ \bibinfo {pages}
  {041307(R)} (\bibinfo {year} {2013})}\BibitemShut {NoStop}%
\bibitem [{\citenamefont {Liao}\ \emph {et~al.}(2015)\citenamefont {Liao},
  \citenamefont {Ou}, \citenamefont {Feng}, \citenamefont {Yang}, \citenamefont
  {Lin}, \citenamefont {Yang}, \citenamefont {Wu}, \citenamefont {He},
  \citenamefont {Ma}, \citenamefont {Xue},\ and\ \citenamefont
  {Li}}]{Liao2015}%
  \BibitemOpen
  \bibfield  {author} {\bibinfo {author} {\bibfnamefont {J.}~\bibnamefont
  {Liao}}, \bibinfo {author} {\bibfnamefont {Y.}~\bibnamefont {Ou}}, \bibinfo
  {author} {\bibfnamefont {X.}~\bibnamefont {Feng}}, \bibinfo {author}
  {\bibfnamefont {S.}~\bibnamefont {Yang}}, \bibinfo {author} {\bibfnamefont
  {C.}~\bibnamefont {Lin}}, \bibinfo {author} {\bibfnamefont {W.}~\bibnamefont
  {Yang}}, \bibinfo {author} {\bibfnamefont {K.}~\bibnamefont {Wu}}, \bibinfo
  {author} {\bibfnamefont {K.}~\bibnamefont {He}}, \bibinfo {author}
  {\bibfnamefont {X.}~\bibnamefont {Ma}}, \bibinfo {author} {\bibfnamefont
  {Q.~K.}\ \bibnamefont {Xue}}, \ and\ \bibinfo {author} {\bibfnamefont
  {Y.}~\bibnamefont {Li}},\ }\href {\doibase 10.1103/PhysRevLett.114.216601}
  {\bibfield  {journal} {\bibinfo  {journal} {Physical Review Letters}\
  }\textbf {\bibinfo {volume} {114}},\ \bibinfo {pages} {216601} (\bibinfo
  {year} {2015})}\BibitemShut {NoStop}%
\bibitem [{\citenamefont {Taskin}\ \emph {et~al.}(2017)\citenamefont {Taskin},
  \citenamefont {Legg}, \citenamefont {Yang}, \citenamefont {Sasaki},
  \citenamefont {Kanai}, \citenamefont {Matsumoto}, \citenamefont {Rosch},\
  and\ \citenamefont {Ando}}]{Taskin2017}%
  \BibitemOpen
  \bibfield  {author} {\bibinfo {author} {\bibfnamefont {A.~A.}\ \bibnamefont
  {Taskin}}, \bibinfo {author} {\bibfnamefont {H.~F.}\ \bibnamefont {Legg}},
  \bibinfo {author} {\bibfnamefont {F.}~\bibnamefont {Yang}}, \bibinfo {author}
  {\bibfnamefont {S.}~\bibnamefont {Sasaki}}, \bibinfo {author} {\bibfnamefont
  {Y.}~\bibnamefont {Kanai}}, \bibinfo {author} {\bibfnamefont
  {K.}~\bibnamefont {Matsumoto}}, \bibinfo {author} {\bibfnamefont
  {A.}~\bibnamefont {Rosch}}, \ and\ \bibinfo {author} {\bibfnamefont
  {Y.}~\bibnamefont {Ando}},\ }\href {\doibase 10.1038/s41467-017-01474-8}
  {\bibfield  {journal} {\bibinfo  {journal} {Nature Communications}\ }\textbf
  {\bibinfo {volume} {8}},\ \bibinfo {pages} {1340} (\bibinfo {year}
  {2017})}\BibitemShut {NoStop}%
\bibitem [{\citenamefont {Xu}\ \emph {et~al.}(2014)\citenamefont {Xu},
  \citenamefont {Miotkowski}, \citenamefont {Liu}, \citenamefont {Tian},
  \citenamefont {Nam}, \citenamefont {Alidoust}, \citenamefont {Hu},
  \citenamefont {Shih}, \citenamefont {Hasan},\ and\ \citenamefont
  {Chen}}]{Xu2014a}%
  \BibitemOpen
  \bibfield  {author} {\bibinfo {author} {\bibfnamefont {Y.}~\bibnamefont
  {Xu}}, \bibinfo {author} {\bibfnamefont {I.}~\bibnamefont {Miotkowski}},
  \bibinfo {author} {\bibfnamefont {C.}~\bibnamefont {Liu}}, \bibinfo {author}
  {\bibfnamefont {J.}~\bibnamefont {Tian}}, \bibinfo {author} {\bibfnamefont
  {H.}~\bibnamefont {Nam}}, \bibinfo {author} {\bibfnamefont {N.}~\bibnamefont
  {Alidoust}}, \bibinfo {author} {\bibfnamefont {J.}~\bibnamefont {Hu}},
  \bibinfo {author} {\bibfnamefont {C.-K.}\ \bibnamefont {Shih}}, \bibinfo
  {author} {\bibfnamefont {M.~Z.}\ \bibnamefont {Hasan}}, \ and\ \bibinfo
  {author} {\bibfnamefont {Y.~P.}\ \bibnamefont {Chen}},\ }\href
  {http://dx.doi.org/10.1038/nphys3140 10.1038/nphys3140
  http://www.nature.com/nphys/journal/v10/n12/abs/nphys3140.html{\#}supplementary-information}
  {\bibfield  {journal} {\bibinfo  {journal} {Nature Physics}\ }\textbf
  {\bibinfo {volume} {10}},\ \bibinfo {pages} {956} (\bibinfo {year}
  {2014})}\BibitemShut {NoStop}%
\bibitem [{\citenamefont {Xu}\ \emph {et~al.}(2016)\citenamefont {Xu},
  \citenamefont {Miotkowski},\ and\ \citenamefont {Chen}}]{Xu2016}%
  \BibitemOpen
  \bibfield  {author} {\bibinfo {author} {\bibfnamefont {Y.}~\bibnamefont
  {Xu}}, \bibinfo {author} {\bibfnamefont {I.}~\bibnamefont {Miotkowski}}, \
  and\ \bibinfo {author} {\bibfnamefont {Y.~P.}\ \bibnamefont {Chen}},\ }\href
  {\doibase 10.1038/ncomms11434} {\bibfield  {journal} {\bibinfo  {journal}
  {Nature Communications}\ }\textbf {\bibinfo {volume} {7}},\ \bibinfo {pages}
  {11434} (\bibinfo {year} {2016})}\BibitemShut {NoStop}%
\bibitem [{\citenamefont {Kim}\ \emph {et~al.}(2012)\citenamefont {Kim},
  \citenamefont {Cho}, \citenamefont {Butch}, \citenamefont {Syers},
  \citenamefont {Kirshenbaum}, \citenamefont {Adam}, \citenamefont {Paglione},\
  and\ \citenamefont {Fuhrer}}]{Kim2012c}%
  \BibitemOpen
  \bibfield  {author} {\bibinfo {author} {\bibfnamefont {D.}~\bibnamefont
  {Kim}}, \bibinfo {author} {\bibfnamefont {S.}~\bibnamefont {Cho}}, \bibinfo
  {author} {\bibfnamefont {N.~P.}\ \bibnamefont {Butch}}, \bibinfo {author}
  {\bibfnamefont {P.}~\bibnamefont {Syers}}, \bibinfo {author} {\bibfnamefont
  {K.}~\bibnamefont {Kirshenbaum}}, \bibinfo {author} {\bibfnamefont
  {S.}~\bibnamefont {Adam}}, \bibinfo {author} {\bibfnamefont {J.}~\bibnamefont
  {Paglione}}, \ and\ \bibinfo {author} {\bibfnamefont {M.~S.}\ \bibnamefont
  {Fuhrer}},\ }\href {\doibase 10.1038/nphys2286} {\bibfield  {journal}
  {\bibinfo  {journal} {Nature Physics}\ }\textbf {\bibinfo {volume} {8}},\
  \bibinfo {pages} {460} (\bibinfo {year} {2012})}\BibitemShut {NoStop}%
\bibitem [{\citenamefont {Fatemi}\ \emph {et~al.}(2014)\citenamefont {Fatemi},
  \citenamefont {Hunt}, \citenamefont {Steinberg}, \citenamefont {Eltinge},
  \citenamefont {Mahmood}, \citenamefont {Butch}, \citenamefont {Watanabe},
  \citenamefont {Taniguchi}, \citenamefont {Gedik}, \citenamefont {Ashoori},\
  and\ \citenamefont {Jarillo-Herrero}}]{Fatemi2014}%
  \BibitemOpen
  \bibfield  {author} {\bibinfo {author} {\bibfnamefont {V.}~\bibnamefont
  {Fatemi}}, \bibinfo {author} {\bibfnamefont {B.}~\bibnamefont {Hunt}},
  \bibinfo {author} {\bibfnamefont {H.}~\bibnamefont {Steinberg}}, \bibinfo
  {author} {\bibfnamefont {S.~L.}\ \bibnamefont {Eltinge}}, \bibinfo {author}
  {\bibfnamefont {F.}~\bibnamefont {Mahmood}}, \bibinfo {author} {\bibfnamefont
  {N.~P.}\ \bibnamefont {Butch}}, \bibinfo {author} {\bibfnamefont
  {K.}~\bibnamefont {Watanabe}}, \bibinfo {author} {\bibfnamefont
  {T.}~\bibnamefont {Taniguchi}}, \bibinfo {author} {\bibfnamefont
  {N.}~\bibnamefont {Gedik}}, \bibinfo {author} {\bibfnamefont {R.~C.}\
  \bibnamefont {Ashoori}}, \ and\ \bibinfo {author} {\bibfnamefont
  {P.}~\bibnamefont {Jarillo-Herrero}},\ }\href {\doibase
  10.1103/PhysRevLett.113.206801} {\bibfield  {journal} {\bibinfo  {journal}
  {Physical Review Letters}\ }\textbf {\bibinfo {volume} {113}},\ \bibinfo
  {pages} {206801} (\bibinfo {year} {2014})}\BibitemShut {NoStop}%
\bibitem [{\citenamefont {{Das Sarma}}\ and\ \citenamefont
  {Hwang}(2014)}]{DasSarma2014}%
  \BibitemOpen
  \bibfield  {author} {\bibinfo {author} {\bibfnamefont {S.}~\bibnamefont {{Das
  Sarma}}}\ and\ \bibinfo {author} {\bibfnamefont {E.~H.}\ \bibnamefont
  {Hwang}},\ }\href {\doibase 10.1103/PhysRevB.89.235423} {\bibfield  {journal}
  {\bibinfo  {journal} {Physical Review B}\ }\textbf {\bibinfo {volume} {89}},\
  \bibinfo {pages} {235423} (\bibinfo {year} {2014})}\BibitemShut {NoStop}%
\bibitem [{\citenamefont {Kim}\ \emph {et~al.}(2013)\citenamefont {Kim},
  \citenamefont {Syers}, \citenamefont {Butch}, \citenamefont {Paglione},\ and\
  \citenamefont {Fuhrer}}]{Kim2013c}%
  \BibitemOpen
  \bibfield  {author} {\bibinfo {author} {\bibfnamefont {D.}~\bibnamefont
  {Kim}}, \bibinfo {author} {\bibfnamefont {P.}~\bibnamefont {Syers}}, \bibinfo
  {author} {\bibfnamefont {N.~P.}\ \bibnamefont {Butch}}, \bibinfo {author}
  {\bibfnamefont {J.}~\bibnamefont {Paglione}}, \ and\ \bibinfo {author}
  {\bibfnamefont {M.~S.}\ \bibnamefont {Fuhrer}},\ }\href {\doibase
  10.1038/ncomms3040} {\bibfield  {journal} {\bibinfo  {journal} {Nature
  Communications}\ }\textbf {\bibinfo {volume} {4}},\ \bibinfo {pages} {2040}
  (\bibinfo {year} {2013})}\BibitemShut {NoStop}%
\bibitem [{\citenamefont {Xiong}\ \emph {et~al.}(2015)\citenamefont {Xiong},
  \citenamefont {Kushwaha}, \citenamefont {Liang}, \citenamefont {Krizan},
  \citenamefont {Hirschberger}, \citenamefont {Wang}, \citenamefont {Cava},\
  and\ \citenamefont {Ong}}]{Xiong2015}%
  \BibitemOpen
  \bibfield  {author} {\bibinfo {author} {\bibfnamefont {J.}~\bibnamefont
  {Xiong}}, \bibinfo {author} {\bibfnamefont {S.~K.}\ \bibnamefont {Kushwaha}},
  \bibinfo {author} {\bibfnamefont {T.}~\bibnamefont {Liang}}, \bibinfo
  {author} {\bibfnamefont {J.~W.}\ \bibnamefont {Krizan}}, \bibinfo {author}
  {\bibfnamefont {M.}~\bibnamefont {Hirschberger}}, \bibinfo {author}
  {\bibfnamefont {W.}~\bibnamefont {Wang}}, \bibinfo {author} {\bibfnamefont
  {R.~J.}\ \bibnamefont {Cava}}, \ and\ \bibinfo {author} {\bibfnamefont
  {N.~P.}\ \bibnamefont {Ong}},\ }\href {\doibase 10.1126/science.aac6089}
  {\bibfield  {journal} {\bibinfo  {journal} {Science}\ }\textbf {\bibinfo
  {volume} {350}},\ \bibinfo {pages} {413} (\bibinfo {year}
  {2015})}\BibitemShut {NoStop}%
\bibitem [{\citenamefont {Goswami}\ \emph {et~al.}(2015)\citenamefont
  {Goswami}, \citenamefont {Pixley},\ and\ \citenamefont {{Das
  Sarma}}}]{Goswami2015}%
  \BibitemOpen
  \bibfield  {author} {\bibinfo {author} {\bibfnamefont {P.}~\bibnamefont
  {Goswami}}, \bibinfo {author} {\bibfnamefont {J.~H.}\ \bibnamefont {Pixley}},
  \ and\ \bibinfo {author} {\bibfnamefont {S.}~\bibnamefont {{Das Sarma}}},\
  }\href {\doibase 10.1103/PhysRevB.92.075205} {\bibfield  {journal} {\bibinfo
  {journal} {Physical Review B}\ }\textbf {\bibinfo {volume} {92}},\ \bibinfo
  {pages} {075205} (\bibinfo {year} {2015})}\BibitemShut {NoStop}%
\bibitem [{\citenamefont {Sivan}\ \emph {et~al.}(1988)\citenamefont {Sivan},
  \citenamefont {Entin-Wohlman},\ and\ \citenamefont {Imry}}]{Sivan1988}%
  \BibitemOpen
  \bibfield  {author} {\bibinfo {author} {\bibfnamefont {U.}~\bibnamefont
  {Sivan}}, \bibinfo {author} {\bibfnamefont {O.}~\bibnamefont
  {Entin-Wohlman}}, \ and\ \bibinfo {author} {\bibfnamefont {Y.}~\bibnamefont
  {Imry}},\ }\href {\doibase 10.1103/PhysRevLett.60.1566} {\bibfield  {journal}
  {\bibinfo  {journal} {Phys. Rev. Lett.}\ }\textbf {\bibinfo {volume} {60}},\
  \bibinfo {pages} {1566} (\bibinfo {year} {1988})}\BibitemShut {NoStop}%
\bibitem [{\citenamefont {Minot}\ \emph {et~al.}(2004)\citenamefont {Minot},
  \citenamefont {Yaish}, \citenamefont {Sazonova},\ and\ \citenamefont
  {McEuen}}]{Minot2004}%
  \BibitemOpen
  \bibfield  {author} {\bibinfo {author} {\bibfnamefont {E.~D.}\ \bibnamefont
  {Minot}}, \bibinfo {author} {\bibfnamefont {Y.}~\bibnamefont {Yaish}},
  \bibinfo {author} {\bibfnamefont {V.}~\bibnamefont {Sazonova}}, \ and\
  \bibinfo {author} {\bibfnamefont {P.~L.}\ \bibnamefont {McEuen}},\ }\href
  {\doibase 10.1038/nature02425} {\bibfield  {journal} {\bibinfo  {journal}
  {Nature}\ }\textbf {\bibinfo {volume} {428}},\ \bibinfo {pages} {536}
  (\bibinfo {year} {2004})}\BibitemShut {NoStop}%
\bibitem [{\citenamefont {Chu}\ \emph {et~al.}(2011)\citenamefont {Chu},
  \citenamefont {Shi},\ and\ \citenamefont {Shen}}]{Chu2011}%
  \BibitemOpen
  \bibfield  {author} {\bibinfo {author} {\bibfnamefont {R.-L.}\ \bibnamefont
  {Chu}}, \bibinfo {author} {\bibfnamefont {J.}~\bibnamefont {Shi}}, \ and\
  \bibinfo {author} {\bibfnamefont {S.-Q.}\ \bibnamefont {Shen}},\ }\href
  {\doibase 10.1103/PhysRevB.84.085312} {\bibfield  {journal} {\bibinfo
  {journal} {Physical Review B}\ }\textbf {\bibinfo {volume} {84}},\ \bibinfo
  {pages} {085312} (\bibinfo {year} {2011})}\BibitemShut {NoStop}%
\bibitem [{\citenamefont {Taskin}\ and\ \citenamefont
  {Ando}(2011)}]{Taskin2011a}%
  \BibitemOpen
  \bibfield  {author} {\bibinfo {author} {\bibfnamefont {A.~A.}\ \bibnamefont
  {Taskin}}\ and\ \bibinfo {author} {\bibfnamefont {Y.}~\bibnamefont {Ando}},\
  }\href {\doibase 10.1103/PhysRevB.84.035301} {\bibfield  {journal} {\bibinfo
  {journal} {Physical Review B}\ }\textbf {\bibinfo {volume} {84}},\ \bibinfo
  {pages} {035301} (\bibinfo {year} {2011})}\BibitemShut {NoStop}%
\bibitem [{\citenamefont {Fu}\ \emph {et~al.}(2016)\citenamefont {Fu},
  \citenamefont {Hanaguri}, \citenamefont {Igarashi}, \citenamefont {Kawamura},
  \citenamefont {Bahramy},\ and\ \citenamefont {Sasagawa}}]{Fu2016}%
  \BibitemOpen
  \bibfield  {author} {\bibinfo {author} {\bibfnamefont {Y.-S.}\ \bibnamefont
  {Fu}}, \bibinfo {author} {\bibfnamefont {T.}~\bibnamefont {Hanaguri}},
  \bibinfo {author} {\bibfnamefont {K.}~\bibnamefont {Igarashi}}, \bibinfo
  {author} {\bibfnamefont {M.}~\bibnamefont {Kawamura}}, \bibinfo {author}
  {\bibfnamefont {M.~S.}\ \bibnamefont {Bahramy}}, \ and\ \bibinfo {author}
  {\bibfnamefont {T.}~\bibnamefont {Sasagawa}},\ }\href {\doibase
  10.1038/ncomms10829} {\bibfield  {journal} {\bibinfo  {journal} {Nature
  Communications}\ }\textbf {\bibinfo {volume} {7}},\ \bibinfo {pages} {10829}
  (\bibinfo {year} {2016})}\BibitemShut {NoStop}%
\bibitem [{\citenamefont {Nam}\ \emph {et~al.}(2017)\citenamefont {Nam},
  \citenamefont {Xu}, \citenamefont {Miotkowski}, \citenamefont {Tian},
  \citenamefont {Chen}, \citenamefont {Liu}, \citenamefont {Hasan},
  \citenamefont {Zhu}, \citenamefont {Fiete},\ and\ \citenamefont
  {Shih}}]{Nam2017}%
  \BibitemOpen
  \bibfield  {author} {\bibinfo {author} {\bibfnamefont {H.}~\bibnamefont
  {Nam}}, \bibinfo {author} {\bibfnamefont {Y.}~\bibnamefont {Xu}}, \bibinfo
  {author} {\bibfnamefont {I.}~\bibnamefont {Miotkowski}}, \bibinfo {author}
  {\bibfnamefont {J.}~\bibnamefont {Tian}}, \bibinfo {author} {\bibfnamefont
  {Y.~P.}\ \bibnamefont {Chen}}, \bibinfo {author} {\bibfnamefont
  {C.}~\bibnamefont {Liu}}, \bibinfo {author} {\bibfnamefont {M.~Z.}\
  \bibnamefont {Hasan}}, \bibinfo {author} {\bibfnamefont {W.}~\bibnamefont
  {Zhu}}, \bibinfo {author} {\bibfnamefont {G.~A.}\ \bibnamefont {Fiete}}, \
  and\ \bibinfo {author} {\bibfnamefont {C.-K.}\ \bibnamefont {Shih}},\ }\href
  {\doibase doi:10.1016/j.jpcs.2017.10.026} {\bibfield  {journal} {\bibinfo
  {journal} {Journal of Physics and Chemistry of Solids}\ } (\bibinfo {year}
  {2017}),\ doi:10.1016/j.jpcs.2017.10.026}\BibitemShut {NoStop}%
\bibitem [{\citenamefont {Takahashi}(2015)}]{Takahashi2015}%
  \BibitemOpen
  \bibfield  {author} {\bibinfo {author} {\bibfnamefont {R.}~\bibnamefont
  {Takahashi}},\ }in\ \href {\doibase 10.1007/978-4-431-55534-6_4} {\emph
  {\bibinfo {booktitle} {Topological States on Interfaces Protected by
  Symmetry}}}\ (\bibinfo  {publisher} {Springer Japan},\ \bibinfo {address}
  {Tokyo},\ \bibinfo {year} {2015})\ pp.\ \bibinfo {pages} {63--71}\BibitemShut
  {NoStop}%
\end{thebibliography}%

\begin{appendix}
\centering
\includegraphics[page=1,scale=0.85]{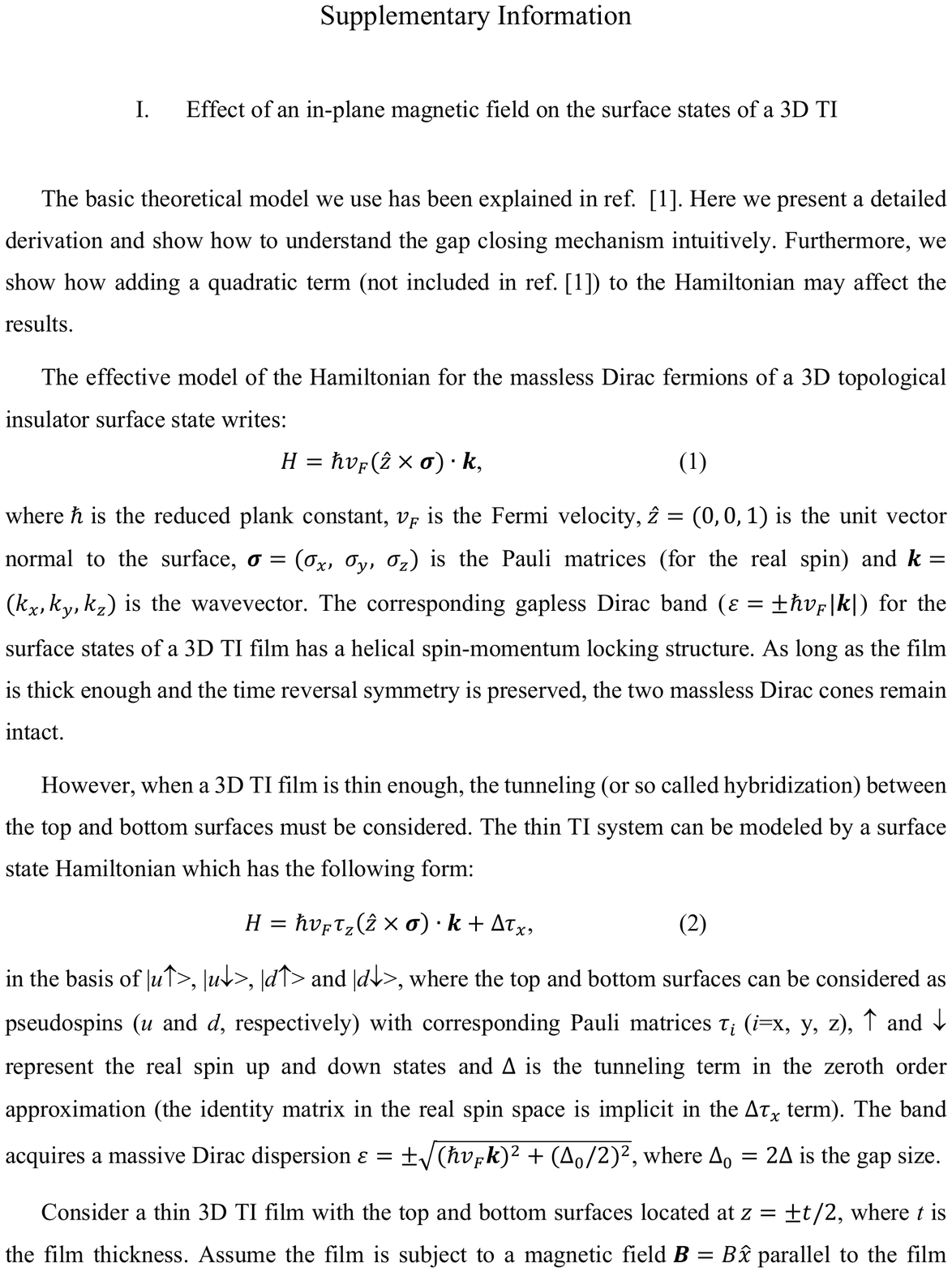} 
\clearpage  
\centering
\includegraphics[page=2,scale=0.85]{Supplementary_information.pdf} 
\clearpage
\centering
\includegraphics[page=3,scale=0.85]{Supplementary_information.pdf} 
\clearpage
\centering
\includegraphics[page=4,scale=0.85]{Supplementary_information.pdf} 
\clearpage
\centering
\includegraphics[page=5,scale=0.85]{Supplementary_information.pdf} 
\clearpage
\centering
\includegraphics[page=6,scale=0.85]{Supplementary_information.pdf} 
\clearpage
\centering
\includegraphics[page=7,scale=0.85]{Supplementary_information.pdf} 
\clearpage
\centering
\includegraphics[page=8,scale=0.85]{Supplementary_information.pdf} 
\clearpage
\centering
\includegraphics[page=9,scale=0.85]{Supplementary_information.pdf} 
\clearpage
\centering
\includegraphics[page=10,scale=0.85]{Supplementary_information.pdf} 
\clearpage
\centering
\includegraphics[page=11,scale=0.85]{Supplementary_information.pdf} 
\clearpage
\centering
\includegraphics[page=12,scale=0.85]{Supplementary_information.pdf} 
\clearpage
\centering
\includegraphics[page=13,scale=0.85]{Supplementary_information.pdf} 
\clearpage	
\end{appendix}

\end{document}